\documentclass[10pt]{iopart}

\usepackage{iopams}  

\usepackage{graphicx}
\usepackage{wrapfig}
\usepackage{hyperref}
\usepackage{xcolor}
\usepackage{cite} 
\usepackage{amssymb}
\usepackage{multirow}
\usepackage{color, colortbl}
\expandafter\let\csname equation*\endcsname\relax     
\expandafter\let\csname endequation*\endcsname\relax  
\usepackage{amsmath}    
\definecolor{Gray}{gray}{0.9}
\newcolumntype{g}{>{\columncolor{Gray}}c}

\definecolor{mygreen}{RGB}{28,172,0} 
\definecolor{mylilas}{RGB}{170,55,241}

\begin{document}

\title{Integrated Data Analysis and Validation} 

\author{R. Fischer$^1$,
A. Bock$^1$,
S. S. Denk$^2$,
A. Medvedeva$^3$,
M. Salewski$^4$,
M. Schneider$^5$,
D. Stieglitz$^1$,
and the ASDEX Upgrade Team$^6$}

\address{$^1$ Max-Planck-Institut f\"{u}r Plasmaphysik, Garching, Germany\\
$^2$ Plasma Science and Fusion Center, MIT, Cambridge, USA\\
$^3$ Aix Marseille University, CNRS, Centrale Marseille, M2P2, Marseille, France\\
$^4$ Department of Physics, Technical University of Denmark, Kgs. Lyngby, Denmark\\
$^5$ ITER Organization, St Paul-lez-Durance Cedex, France\\
$^6$ See author list of U. Stroth et al. 2022 Nucl. Fusion 62 042006}
\ead{Rainer.Fischer@ipp.mpg.de}


\begin{abstract}
A major challenge in nuclear fusion research is the coherent combination of data from heterogeneous diagnostics and modelling codes for machine control and safety as well as physics studies. Measured data from different diagnostics often provide information about the same subset of physical parameters.  Additionally, information provided by some diagnostics might be needed for the analysis of other diagnostics.  A joint analysis of complementary and redundant data allows, e.g., to improve the reliability of parameter estimation, to increase the spatial and temporal resolution of profiles, to obtain synergistic effects, to consider diagnostics interdependencies and to find and resolve data inconsistencies.  Physics-based modelling and parameter relationships provide additional information improving the treatment of ill-posed inversion problems.  A coherent combination of all kind of available information within a probabilistic framework allows for improved data analysis results.  

The concept of Integrated Data Analysis (IDA) in the framework of Bayesian probability theory is outlined and contrasted with conventional data analysis.
Components of the probabilistic approach are summarized and specific ingredients beneficial for data analysis at fusion devices are discussed. 
\end{abstract}

\maketitle

\ioptwocol

\section{Introduction}

\label{sec-intro}

In present and future fusion devices huge amounts of measurements coming from many diagnostic systems have to be analyzed. 
The information obtained from these measurements are and will be used for machine control and safety as well as for physics studies. 
The goal of the Integrated Data Analysis (IDA) method is to integrate measured data and their analyses to optimize information available for plasma operation and physics studies.
The measured data from diagnostics providing redundant or complementary information are combined, together with available physics knowledge and modelling information within a probabilistic framework.

The Integrated Data Analysis and Validation specialist working group within the International Tokamak Physics Activity (ITPA) Diagnostics Topical Group was founded in the year 2020.
It was motivated by the usefulness of IDA applications at present day machines \cite{FISCHER03b,DODT08b,FORD09a,DODT09a,FISCHER10a,RATHGEBER10a,FISCHER11a,REUSCH14a,GALANTE15a,Salewski2018b,FISCHER20a,KWAK21a,KWAK21b}.
The goal of the Integrated Data Analysis and Validation specialist working group is to provide and apply an IDA framework for present and next generation fusion devices such as ITER and DEMO for self-consistent data analysis and validation procedures.

A comparison of the concept of IDA with a traditional approach for data analysis, based on the analysis of individual diagnostics data and a subsequent combination of the results, can be found in \cite{FISCHER07a}.

IDA in the framework of Bayesian probability theory provides a concept to analyse a coherent combination of measured data from heterogeneous diagnostics and to combine them with physics knowledge and modelling information \cite{FISCHER04b}. 
Since every piece of information from measurements and modelling are subject to uncertainties, quantification and processing of uncertain information is central to this probabilistic approach. 
Complex error propagation is obtained automatically combining data in a concise probabilistic one-step analysis.
The extended set of measurements and modelling information allows for an improved treatment of ill-posed inversion problems of, e.g., profile reconstruction, tomography or equilibrium reconstruction. 
Different techniques for measuring the same subset of physical parameters provide complementary and redundant data for, e.g., improving the reliability of physical parameters, increasing the spatial and temporal resolution of profiles, resolving data inconsistencies, and for reducing the ambiguity of parameters to be estimated without employing non-physical constraints. 

IDA was developed and first applied to reconstruct electron density $n_\textnormal{e}$ and temperature $T_\textnormal{e}$ profiles from a probabilistic analysis of Thomson scattering (TS) data \cite{FISCHER02a} in combination with interferometry and soft X-ray measurements at the W7-AS stellarator \cite{FISCHER03b}. 
A corresponding application at the ASDEX Upgrade tokamak includes additionally to the TS and interferometry data also data from electron cyclotron emission (ECE) and the lithium beam (LIB) diagnostic for which an improved forward model was developed \cite{FISCHER08a,FISCHER10a}.
This LIB forward model was additionally used in a probabilistic analysis of the JET LIB diagnostic \cite{DODT09a}.
At W7-AS, Bayesian graphical models were introduced for integrating diagnostic data analyses \cite{SVENSSON04a}.
IDA was then applied at ASDEX Upgrade to reconstruct the effective ion charge $Z_\textnormal{eff}$ profiles from various charge exchange recombination spectroscopy (CXRS) measurements \cite{RATHGEBER10a}.
A non-Maxwellian electron energy distribution function in the positive column of a neon dc-discharge was reconstructed from the visible emission spectrum using IDA \cite{DODT08b}.
At JET the Bayesian combined analysis of LIDAR, edge LIDAR and
interferometry diagnostics provided $n_\textnormal{e}$ and $T_\textnormal{e}$ profiles \cite{FORD09a}.
At the TJ-II stellarator the $n_\textnormal{e}$ profile was reconstructed in an IDA approach using data from interferometry, reflectometry, TS, and the helium beam \cite{FISCHER11a}.
At the Madison Symmetric Torus (MST) reversed field pinch (RFP) the $T_\textnormal{e}$ profiles were estimated in the probabilistic framework from a combination of the double-foil soft
X-ray system (SXR) and the TS diagnostic \cite{REUSCH14a}.
Additionally, at the MST RFP $Z_\textnormal{eff}$ profiles were determined by the integration of soft X-ray tomography and charge exchange recombination spectroscopy impurity density measurements \cite{GALANTE15a}.
The ion temperature $T_\textnormal{i}$ and rotation profiles $v_\textnormal{rot}$ were reconstructed at ASDEX Upgrade in a probabilistic integrated approach from various charge exchange recombination spectroscopy (CXRS) measurements using Gaussian process regression \cite{FISCHER20a}.
Recent Bayesian analyses combining various diagnostics can be found at W7-X \cite{KWAK21a,KWAK21b}, at ASDEX Upgrade and JET \cite{Salewski2018b}, and the MST RFP \cite{REUSCH18a}.

The present paper aims at showing the basic ingredients of Integrated Data Analysis in the Bayesian framework, reviewing the work previously done, and some examples highlighting typical realizations. 
More details of the implementation of IDA can be found in publications which are cited as appropriate.
The references in this paper can not be exhaustive as they should only provide an entrance point for the interested reader.

Section \ref{sec-IDA} compares IDA with a conventional data analysis approach for multiple-diagnostic data analysis and summarizes the main ingredients of IDA: Bayesian probability theory (section \ref{sec-bayes-PT}), forward models (\ref{sec-fw-model}), uncertainty quantification (\ref{sec-unc-quantification}), likelihoods (\ref{sec-likelihoods}), prior information (\ref{sec-priors}), parameterization (\ref{sec-parameterisation}), methods for parameter and uncertainty estimation (\ref{sec-params-estimation}), validation (\ref{sec-validation}) and numerical implementation (\ref{sec-num-impl}). 
Section \ref{sec-imas} addresses the ITER Integrated Modelling \& Analysis Suite (IMAS) for physics modelling and data analysis as a standardized way to access and process data.
Section \ref{sec-example} shows examples applying IDA to obtain synergistic effects (\ref{sec-synergy}), profile reconstruction (\ref{sec-prof-recon}), equilibrium reconstruction (\ref{sec-equi-recon}) and velocity-space tomography (\ref{sec-tomo}).
Section \ref{sec-summary} summarises.

\section{Integrated Data Analysis}
\label{sec-IDA}

\begin{figure}[tbp!]
\centering
\includegraphics[clip,width=80mm]{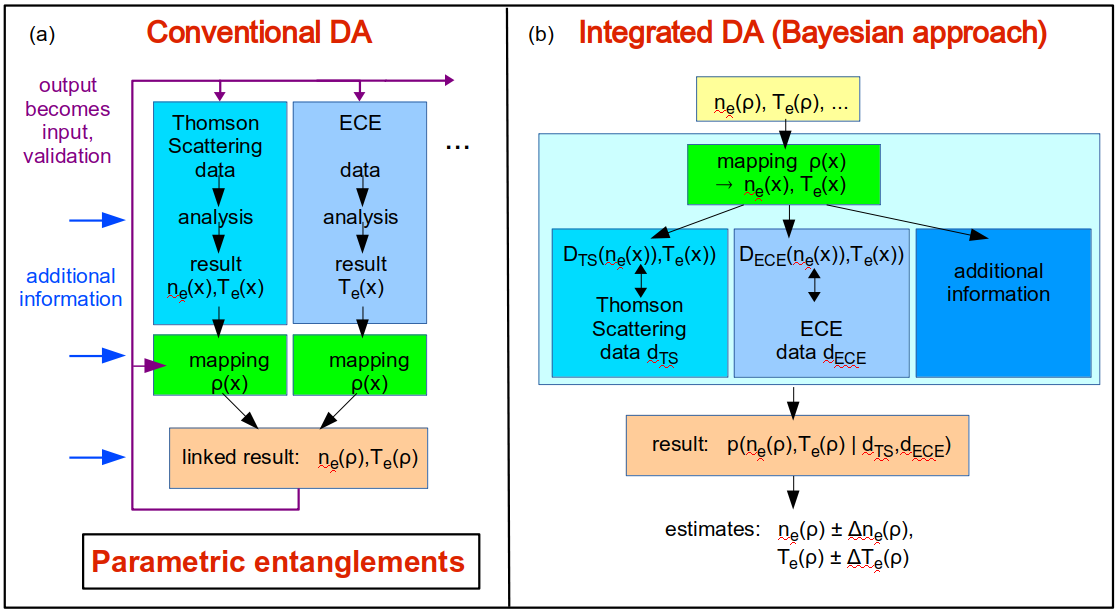}
\caption{Simplified flow-charts for typical data analysis steps inferring electron temperature $T_\textnormal{e}$ and
density $n_\textnormal{e}$ profiles for magnetic confinement fusion experiments from the Thomson scattering and electron
cyclotron emission (ECE) diagnostics in (a) a conventional approach and (b) within the IDA concept.}
\label{fig:ida_comp}
\end{figure}

The IDA approach in the framework of Bayesian probability theory is conceptually different from an often used sequential ({\em conventional}) data analysis approach.
Frequently, due to the large amount of diagnostics available at fusion devices, in conventional data analysis the individual diagnostics are analysed by the responsible diagnosticians familiar with the hardware, physics and analysis details (Fig.\ \ref{fig:ida_comp}(a)).
To obtain a unique ({\em linked}) result the various results of the heterogeneous diagnostics are mapped on a common, typically magnetic, coordinate system and fitted with a joint parameter set.
Often the analysis of the single diagnostics are augmented with additional information to regularize ill-posed inversion problems and obtain, e.g., smooth and well-defined results. 
The linked result might then be used as input for the equilibrium (mapping) estimation, the analysis of other diagnostics or for result validation and consistency checks.

Various challenges of this conventional approach arise from the parametric entanglements involved. 
In this iterative procedure it might be difficult to obtain a (self-)consistent result, in particular if many diagnostics are involved as is the case for present and future fusion devices.
An automation of this analysis chain is challenging if a huge amount of data has to be analyzed.
The propagation of information between diagnostics might be incomplete if single estimates from one diagnostic are used as input for other diagnostics, neglecting complex parameter interdependencies.
Error propagation is frequently neglected, resulting in an underestimation of the estimation uncertainties.
Data and result validation and overall consistency checks between coupled diagnostics might be a non-trivial task because a quantitative and unified description and processing of statistical and systematic uncertainties is missing.
Furthermore, often backward inversion techniques are used which might be prone to noise fitting or numerical instabilities necessary to be regularized by additional constraints or data binning.
The estimated parameters (linked result) and their uncertainties often miss a description of the non-linear dependencies.

These difficulties are resolved by the IDA approach using a probabilistic combination of different diagnostics (Fig.\ \ref{fig:ida_comp}(b)).
The scheme starts with a complete set of physical parameters (section \ref{sec-parameterisation}), as a function of a common coordinate system, which is sufficient to describe all diagnostics data.
Only forward modelling is used which allows one to evaluate the diagnostics data given the parameters of interest (\ref{sec-fw-model}). 
Forward modelling is known to be numerically stable.
The measured data of a diagnostic is compared to the forward modelled data with a likelihood probability distribution function (pdf) (\ref{sec-likelihoods}) describing the distribution of the data uncertainties  (\ref{sec-unc-quantification}). 
Additional physical information and its uncertainty can readily be integrated with a probabilistic description and is used only once in the analysis of the combined set of diagnostics (\ref{sec-priors}). 
Systematic effects are described with {\em nuisance} parameters. Their uncertainties are quantified with probability distributions.
The nuisance parameters are integrated out ({\em marginalized}) such that the uncertainty in these parameters propagate to the uncertainty of the parameters of interest.
Uncertain nuisance parameters can arise from calibration {\em constants}, atomic data, or quantification of systematic effects which are candidates for diagnostic inconsistencies.
Quantification of inconsistency effects might help to resolve the reason for diverging diagnostics results.
Finally, the result of the Bayesian approach is a multi-dimensional probability distribution which quantifies how reliable a certain set of parameter values is in the light of all measured data and the additional information provided (\ref{sec-bayes-PT}).
This {\em posterior} probability distribution includes all parameter interdependencies.
Low-dimensional properties of this posterior pdf allows for estimating the parameters (maximum or mean), their dependencies (covariance) and their uncertainties (variance) (\ref{sec-params-estimation}).

\subsection{Bayesian probability theory}
\label{sec-bayes-PT}

The interpretative and numerical framework of Integrated Data Analysis is given by Bayesian probability theory (BPT).
BPT provides a unique interpretation of probability as a measure of uncertainty and rules to combine and process (uncertain) information.
Uncertainty in the Bayesian framework is lack of knowledge.
An introduction to Bayesian inference and further references can be found in \cite{TOUSSAINT11a}.
Measured data as well as most information used to describe the measured data or constrain the parameter space of interest suffer from uncertainties.
Therefore, a unique framework to handle any kind of uncertainty is mandatory if different sources of information have to be analysed jointly.
An overview of various types of uncertainties encompassed by BPT is given in section \ref{sec-unc-quantification}.

Additional to the unique quantification of information and its uncertainty, BPT provides rules to combine and process information.
Bayes' theorem in its most reduced form relates the posterior pdf $p(f|d)$ for the parameters of interest $f$ given the data $d$ with the likelihood pdf $p(d|f)$, the prior pdf $p(f)$ and the evidence $p(d)$:
\begin{eqnarray}
p(f|d) = \frac{p(d|f) \times p(f)}{p(d)}.
\label{eq:bayes}
\end{eqnarray}
$p(A|B)$ means probability that $A$ is true given (assuming) $B$ is true.
The power of Bayes' formula is that it provides what we actually want to know from what the forward model by themselves can produce. The likelihood pdf with the forward model of the measured data provide the probability of the measurement data, given the parameters of interest, $p(d|f)$. 
But we actually want to know the probability (reliability) of the parameters of interest, given the diagnostic data, $p(f|d)$.

The Bayesian scheme can be expanded by taking the product of all likelihood pdfs describing all diagnostic measurements and prior pdfs describing any kind of additional physical information used. 
Therefore, the likelihood $p(d|f)$ in the IDA framework consists of the product of the likelihoods of the various diagnostic data to be analysed jointly,
$p(d|f) = \prod_k p_k(d_k|f)$, where the data of diagnostic $k$, $d_k$, are described with the likelihood $p_k(d_k|f)$.
The functional form of a likelihood pdf depends on the uncertainty distribution of the measured data and might differ for the various diagnostics (see section \ref{sec-unc-data}).
As a likelihood pdf describes the probability of measuring a certain data value assuming one knows the underlying physics, the likelihood links the measured data with a forward model (FM) of the measurement process $D_k(f)$ (\ref{sec-fw-model}), which typically also varies for the heterogeneous diagnostics.
The likelihood for each diagnostic, $p_k(d_k|f)$, is typically a product of likelihoods of the measured data points if the data uncertainties are uncorrelated or a multi-dimensional likelihood if the uncertainties are correlated. 

The prior pdf, $p(f)$, describes what we know about the parameter of interest independent of the measurements.
Typical information to be encoded in the prior are non-negativity constraints for, e.g., temperature and density, monotonicity constraints, smoothness constraints, constraints from physics modelling such as for profile gradients.
More examples for useful prior information can be found in the velocity-space tomography section \ref{subsec-prior-tomo}.

\subsection{Forward models}
\label{sec-fw-model}

A forward model (FM) evaluates synthetic data to be compared with the measured data within the likelihood pdf.
Various fidelity levels of FMs for a diagnostic might be available for various purposes.
High-fidelity FMs are typically used for post-plasma analysis where the most reliable results are aimed at. 
For post-plasma analyses numerical resources and time restrictions are less crucial.
Low-fidelity FMs are typically used for time critical applications as real-time analyses or if numerical resources are limited.
An example for a low-fidelity FM is given by ECE analyses where optically thick thermal plasmas are assumed.
For optically thick plasmas black-body radiation can be assumed which results in a radiation temperature which equals the electron temperature, $T_{\rm rad}=T_{\rm e}$.
This ECE FM belongs to one of the simplest FMs where the parameter of interest, here $T_{\rm e}$, is proportional to the measured intensity.
A high-fidelity ECE FM is given by solving the radiation transport equation \cite{DENK18a,DENK20a}.
An implementation of the radiation transport FM can be found in the ECRad code \cite{DENK20a}. 
The ECRad FM is capable of analysing optically thin plasmas with broadened EC emission regions due to high temperatures, as expected for ITER, or due to low-density scenarios.
Additionally, it is capable to describe oblique ECE measurements, harmonic overlap, different polarizations and emission from non-thermal electron energy distribution functions \cite{DENK21a}.
For the analysis of ECE data with this sophisticated model electron density $n_{\rm e}$ profiles are necessary.
Therefore, a combination of the ECE diagnostic with density diagnostics, e.g. Thomson scattering or interferometry, within an IDA framework is mandatory. 

Another example for a multiple-fidelity FM is given by charge exchange recombination spectroscopy (CXRS).
Frequently, ion quantities as ion temperature $T_{\rm i}$ and rotation velocity are pre-evaluated and available from databases.
These data allow for low-fidelity FMs as the parameter of interest, a profile of the ion parameter, can easily be evaluated at the measurement positions.
Ideally, the database provides also information about the uncertainties of the measured values.
A high-fidelity FM for CXRS data describes directly the measured spectra \cite{McDermott17a}.
Although numerically more expensive, this high-fidelity FM would allow to incorporate nuisance parameters describing, e.g., uncertainties in calibration quantities or in the atomic data \cite{DODT08b}.

The preparation of a forward model and its combination with other diagnostics forward models in a probabilistic framework is typically less challenging than developing and combining inversion techniques which additionally might suffer from noise fitting or numerical instabilities as for, e.g., Abel inversion techniques. An example of a    multiple-purpose forward model for velocity-space tomography ({\em weight function}) can be found in Section \ref{sec-tomo}.

\subsection{Uncertainty quantification}
\label{sec-unc-quantification}

Uncertainties in the Bayesian framework are interpreted as lack of knowledge covering any type of uncertainty.
Statistical uncertainties are distinguished from systematic
uncertainties which, in contrast to statistical uncertainties, cannot be reduced by increasing the data sample.
As the results of data analysis depend critically on the uncertainties associated with the data, 
the quantification of uncertainties of measured data (likelihood) and of additional information (prior) is a major part of a parameter estimation problem.
Uncertainties determine the absolute amount of information available and determine the weight of measurements of various diagnostics and prior information relative to each other.
Especially for the detection and resolution of inconsistent measurements, uncertainties play a major role as consistency is obtained if all data and prior information are {\em reasonably} well described within their uncertainties.
Details about the interpretation and definition of uncertainties can be found in {\em Evaluation of measurement data – Guide to the expression of uncertainty in measurement (GUM)} \cite{GUM08a}.

\subsubsection{Uncertainties in measured data}
\label{sec-unc-data}

Statistical measurement noise is always quantified with the likelihood pdf (section \ref{sec-likelihoods}).
A systematic measurement uncertainty is typically described with a prior pdf (\ref{sec-unc-model}). For example, a calibration uncertainty can be quantified with a prior pdf on a calibration nuisance parameter.
In special cases this systematic uncertainty can be propagated to the likelihood pdf (see (\ref{sec-likelihoods}) and  \cite{FISCHER02a}).

The distribution of measurement noise is frequently described with a standard deviation. 
Higher moments are often neglected. 
This defines the use of a multivariate Gaussian distribution suitable for normally distributed measurement errors.
Depending on the measurement scheme other distributions such as the Poisson distribution for counting measurements might be suitable. 
The measurement uncertainty of the lithium beam diagnostic at ASDEX Upgrade was estimated by assuming a Poisson distribution for the photon counts with unknown offset and amplification factor of the measured signal \cite{FISCHER08a}.
In case of unknown measurement uncertainties, contributions to the measurement with unknown source or contributions not described in the forward model, or in case of data failure, robust estimation techniques are mandatory as described in (\ref{sec-likelihoods}).

\subsubsection{Uncertainties in physics models}
\label{sec-unc-model}

Forward models describing measured data or physical models providing prior knowledge to constrain the parameter space frequently are not exactly known and are, therefore, subject to uncertainties.
Typical uncertainties arise from uncertainties in calibration "constants" from calibration measurements, from degrading effect of, e.g., optical components or glas fibers, or from atomic data which themselves are determined by measurements or uncertain modelling.
An example for this is given in section \ref{subsec-unc-tomo} for uncertainties in the forward model (weight matrix) for the velocity-space tomography.

Such systematic uncertainties are tackled in the Bayesian framework by  {\em nuisance} parameters which describe the variability of the model due to the unknown systematic effect. 
The uncertainty of the nuisance parameters are quantified with prior distributions and marginalised (integrated out).
This way the uncertainty of the nuisance parameter propagates into the uncertainty of the parameter of interest.
A systematic ({\em bias}) uncertainty might arise due to uncertainties in the prior information as, e.g., the functional form and weight of the regularization term (see Section \ref{subsec-unc-tomo}).

\subsubsection{Uncertainties in estimated quantities}
\label{sec-unc-params}

Uncertainties in estimated quantities should describe the reliability with which parameters can be inferred from measured data and modelling information. 
These uncertainties arise due to statistical and systematic measurement uncertainties, uncertainties in the forward model and uncertainties in the prior information used.
Various methods for uncertainty estimation exist which are summarized in section [\ref{sec-params-estimation}]. 

\subsection{Likelihoods}
\label{sec-likelihoods}

The likelihood pdf quantifies the probability of measuring a certain data set given the forward model which links the parameters of interest and the measured quantity.
Since the probability of measuring certain data is closely related to the measurement uncertainties, 
the likelihood quantifies the uncertainty distribution of the data.
The most often used likelihood is given by the Gaussian pdf 
with the familiar $\chi^2$-misfit between the data $d$ and the forward modelled data $D(f)$
\begin{eqnarray}
p(d|f) &\propto& \exp\{-\chi^2/2\}  \nonumber \\
\chi^2 &=& \sum_i \chi_i^2 = \sum_i (d_i-D_i)^2 / \sigma_\textnormal{i}^2
\label{eq:gauss}
\end{eqnarray}
given here in its most simplified version.
The use of the Gaussian likelihood is justified for normally distributed measurement errors $\epsilon$ with variance $\langle \epsilon^2 \rangle = \sigma^2$ and mean error $\langle \epsilon \rangle = 0$.
Frequently only the measurement uncertainty describing the statistical distribution of the measurement error is considered. 
The Bayesian interpretation of measurement uncertainties additionally comprises systematic uncertainties from, e.g., calibration or modelling uncertainties 
which can be considered in the likelihood pdf in special cases. 
An example of the use of the Gaussian likelihood with an extended interpretation of the uncertainty can be found in the analysis of Thomson scattering data measured at the stellarator  Wendelstein 7-AS \cite{FISCHER02a}.
The statistical uncertainties of the TS data are augmented with the uncertainties of the background-estimation data, the uncertainty of the calibration measurement, uncertainties of physical model parameters and uncertainties of measured nuisance parameters.
A sensitivity analysis of the uncertainties and model parameters allows for finding the crucial uncertainties which have most impact on the diagnostic performance \cite{FISCHER02a}.

\begin{figure}[tbp!]
\centering
\includegraphics[clip,width=80mm]{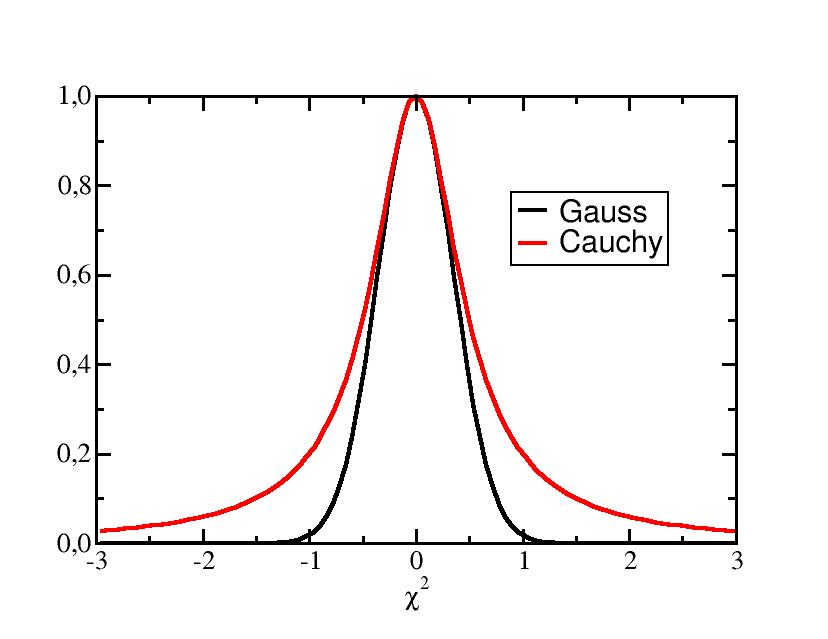}
\caption{Comparison of a Gaussian with a Cauchy distribution appropriate for outlier robust estimation}
\label{fig:gauss_cauchy}
\end{figure}

If measurements suffer from outliers, e.g., due to mis-specified uncertainties, measurement failure or physical contributions not included in the forward model, an outlier robust likelihood is recommended.
The Student's t-distribution treats outliers leniently due to its heavy tails \cite{DOSE99a}
\begin{eqnarray}
p(d|f) &\propto& \prod_i \{a + \chi_i^2 / 2 \}^{-(a+1/2)} \; .
\label{eq:cauchy}
\end{eqnarray}
The Cauchy pdf is obtained for $a=1/2$ and the Gaussian pdf in the limit of $a \rightarrow \infty$.
The heavy tails give outlying data less weight in the fitting process than the Gaussian pdf (Fig.\ \ref{fig:gauss_cauchy}).
The Student's t-distribution can also be used if the standard deviation of the uncertainty is not known \cite{DOSE99a}.
The parameter $a$ of the Student's t-distribution allows one to select the weights of the wing, and therefore the weight outlying data have in the fitting process.

This outlier robust likelihood is used routinely at ASDEX Upgrade for estimating electron temperature and density profiles in the IDA framework \cite{FISCHER10a}.
Examples for outlying data are given in the following examples:
Fringe jumps in interferometry measurements, remaining after a fringe-jump correction procedure, 
typically occur for rapid density changes due to, e.g., pellet injection or signal cross-over due to ion cyclotron resonance heating (ICRH).
Electron cyclotron emission (ECE) data might be deteriorated, e.g., from cut-off, non-thermal electron distributions or harmonic overlay when not described properly by the standard black-body radiation assumption or the more sophisticated radiation transport modelling \cite{RATHGEBER13a,DENK18a,DENK20a,DENK21a}. 
Thomson scattering data might be affected by non-Gaussian calibration uncertainties, low signal-to-noise ratio especially at the low-density edge of the plasma, or by transient events such as filaments which are resolved in the TS diagnostic, which typically has a temporal resolution of about 20~ns, and which are not resolved with other diagnostics.
Lithium beam data might be deteriorated by beam drifts typically not covered by the calibration procedure performed after a plasma discharge, local filaments not measured simultaneously at the positions of the interferometry or TS channels, or background subtraction uncertainties due to frequent events (ELMs) during the beam-off phases \cite{FISCHER08a}.
The emission profile of the thermal helium beam data is typically located at the plasma edge with low signal-to-noise ratio (SNR) at the far scrape-off-layer with low density and temperature values and within the separatrix where the neutral helium beam diminishes due to ionization \cite{WENDLER22a}. 
This low SNR in the intensities easily produces outlying data in the line-ratio data when the denominator comes close to zero destroying any Gaussian assumption about the error distribution.

\subsection{Prior information}
\label{sec-priors}

The Bayesian approach allows to combine measured data from multiple diagnostics with additional information from physical considerations.
In the Bayesian terminology the data independent information and its uncertainty/reliability is quantified with the prior pdfs.

Ubiquitous in profile or tomographic reconstruction is the assumptions of some degree of smoothness, non-negativity or monotonicity.
Smoothness constraints are typically applied using Tikhonov regularization. 
Most often Tikhonov regularization is used to penalize the amplitude (zeroth-order), gradient (first-oder) or curvature (second-order) of distributions. An example for Tikhonov regularization can be found in Section \ref{sec-tomo} with velocity-space tomography.

Non-negativity constraints are less frequently applied due to its degrading performance in the optimization steps.
Nevertheless, optimization routines for estimating best fitting parameters providing boundary constraints are available although at the expense of increased computation time.
An example for the use of a non-negativity constraint can be found in the tomographic reconstruction example in Section \ref{sec-tomo}.
An alternative for boundary constraints for parameters is to quantify a positive parameter with an exponential, e.g., $T(x)=\exp(f(x))$ where $f(x)$ could be an unbounded spline representation.

The exponential of a spline is used at ASDEX Upgrade for the estimation of the temperature and density profiles within the IDA framework and the estimation of the effective ion charge $Z_\textnormal{eff}=1+\exp(f(x))$ which lower bound is 1 \cite{RATHGEBER10a}.
An unbounded estimation of $Z_\textnormal{eff}$ can easily go below 1 due to uncertainties in the data but also due to a deteriorating calibration of the data in case of, e.g., degrading optical components. 
The exponential representation of $Z_\textnormal{eff}$ avoids values below one which is justified if the data are described reasonably well within their uncertainties. 
As an estimation of $Z_\textnormal{eff}=1$ is physically meaningful, it is frequently an indication of a problematic data set.
At ASDEX Upgrade $Z_\textnormal{eff}$ is estimated from the line-integrated bremsstrahlung background of charge-exchange recombination spectra (CXRS) \cite{RATHGEBER10a}.
A degradation of an optical component (coatings on lenses or mirrors, degradation of glass fibers) of the CXRS system is most sensitively detected with an estimated $Z_\textnormal{eff}$ value at the lower limit for a clean, high-density discharge closely after boronization where $Z_\textnormal{eff}$ is expected to be between 1.0 and 1.2.
If, additionally to an estimated $Z_\textnormal{eff}=1$, the residuals between the measured and forward modelled bremsstrahlung data is systematically negative, this is a clear indication of a degradation of an optical component.
This $Z_\textnormal{eff}$ criterion is more sensitive to a deterioration of the calibration than monitoring impurity concentrations determined by CXRS.
Future fusion devices working in a harsh environment might use the $Z_\textnormal{eff}=1$ criterion together with the data residuals for an early detection and quantification of a degradation of optical components and to specify the need for a re-calibration or a cleaning procedure.

Monotonicity constraints or penalty for non-monotonicity can easily be applied similarly to Tikhonov regularization 
\begin{eqnarray}
p(f) &\propto& \exp\{-\sum_i (df(x_i)/dx)^2/(2 \sigma_\textnormal{m}^2)\}
\label{eq:monotonicity}
\end{eqnarray}
where the sum goes over positions $x_i$ where $df(x_i)/dx$ has the wrong sign.
$\sigma_\textnormal{m}$ quantifies the amount of tolerance from the monotonicity penalty. As $\sigma_\textnormal{m}$ decreases, the penalty becomes a constraint.
A strongly monotonic function can be obtained equivalently to the non-negativity constraint by using for the function derivative an exponential of, e.g., a spline. A subsequent integration with appropriate boundary conditions then yields a monotonic function without applying a non-monotonicity penalizing prior. 

Another valuable prior information might arise from physical modelling.
Examples are discussed in section \ref{sec-tomo} with the example of velocity-space tomography where, e.g., the velocity space is restricted or penalized due to simulations or the slowing-down physics is included as a regularizing prior.

\subsection{Parameterization}
\label{sec-parameterisation}

The assignment of the parameter space affects the results of data interpretation.
As shown in section \ref{sec-priors} the choice of parameters allows one to include physics knowledge as positivity, boundary or monotonicity constraints via parameter space reduction.
Additionally, the choice of the parameter set determines the flexibility of the results.
For example, the number and position of spline knots in profile reconstruction determines the spatial resolution.
A reduction of the number of spline knots as well as an increase of the spline knot distance reduces the spatial resolution of data fitting.
Similarly to the smoothness priors, a sparse parameter setting is suitable to reduce noise fitting as well as to mitigate fitting of problematic data if an outlier robust likelihood pdf is applied.

Comparable to the flexibility in the number of the parameters, Gaussian process regression (GPR) allows to reduce profile flexibility by introducing spatial correlation. 
Gaussian process regression is applied, e.g., in fits to electron density and temperature profile data and the estimation of impurity transport coefficients from Alcator C-Mod \cite{CHILENSKI15a}, in the reconstruction of various plasma parameters as in the estimation of ion temperature and rotation profiles at ASDEX Upgrade \cite{FISCHER20a}, for estimating $Z_\textnormal{eff}$ profiles from line integrated bremsstrahlung spectra at Wendelstein 7-X \cite{KWAK21a}, and tomography for soft x-ray spectroscopy at WEST \cite{WANG18a}.
GPR is beneficial for linear problems, e.g., for interpolation and smoothing of noisy data. 
For these cases GPR is computationally fast because analytical formulas for the best estimate and for the estimation uncertainty are available.
Additionally, with a Monte Carlo sampling approach any derived profile, e.g., logarithmic profile gradient and its uncertainties \cite{FISCHER20a}, can efficiently be calculated using the covariance matrix. 

To estimate how much flexibility in the parameter setting is needed, e.g.\ how much spline knots to be chosen for the profiles, criteria are necessary for complexity estimation.
The preferred criterion is to allow as much flexibility as necessary to describe the significant information in the data and reduce flexibility otherwise to avoid noise fitting (Occam's Razor).
Various Bayesian and non-Bayesian techniques are available to (automatically) select the necessary flexibility \cite{FISCHER00a,TOUSSAINT11a,CHILENSKI15a}.

\subsection{Methods for parameter and uncertainty estimation}
\label{sec-params-estimation}

The result of a Bayesian analysis is a {\em posterior} pdf for the parameters of interest given all the data and additional information.
The posterior pdf quantifies how reliable a set of parameters is in the light of the information used.
It contains all the parameter interdependencies.

Parameter estimates can be obtained with various methods distinguishing different properties of the posterior pdf. 
The most popular estimate is given by the maximum-a-posterior (MAP) solution where the posteriori pdf is maximized with respect to the parameters.
For numerical reasons it is preferred to maximize the logarithm of the pdf. 
For Gaussian likelihood and prior pdfs, this is equivalent to minimizing the sum of all $\chi^2$-terms.
The uncertainty of the estimate can be derived from the covariance matrix of the parameters at the MAP solution.
This is equivalent to approximating the typically non-Gaussian posterior pdf with a Gaussian pdf at the MAP estimate ({\em Laplace approximation}) \cite{TOUSSAINT11a}.
The parameter covariance includes the parameter dependencies but fails for strongly asymmetric pdfs as they occur, e.g., when non-negativity constraints are applied. 

For an alternative estimate the mean of the posterior pdf can be used which is different from the MAP estimate for asymmetric pdfs.
Asymmetric pdfs typically occur for non-linear forward models.
Since usually the mean of a multidimensional pdf with non-linear parameter dependencies are not available analytically, Monte-Carlo (MC) sampling methods are used to explore the full pdf \cite{TOUSSAINT11a}.
Among various sampling methods, Markov chain Monte Carlo (MCMC) sampling is frequently used because it is efficient and rather easy to implement.
It allows to sample the full parameter space, to find multimodal pdfs with multiple estimate candidates, to visualize marginal distributions for finding an unresolved subspace or parameter correlations not resolved by the data.
In case of an uni-modal posterior pdf, the mean of the parameter samples provide an estimate, which can be compared to the MAP solution, and the covariance of the samples provides information about the estimation uncertainty.
Furthermore, any derived quantity from a parameter sample can also be averaged and its uncertainty estimated.
As an example see \cite{FISCHER20a} where the electron density and temperature profiles were estimated from MCMC sampling of the posterior pdf of an IDA approach of multiple diagnostics at ASDEX Upgrade.
The uncertainties of the profiles as well as estimates for the profile gradient and the logarithmic gradients and their uncertainties were obtained applying MCMC sampling.

In any case when a new data inference problem is tackled, it is recommended to explore a posterior pdf with MCMC methods at least once to learn about the subtleties of the data analysis problem at hand.


\subsection{Validation}
\label{sec-validation}

The validation of the results from a Bayesian analysis is closely related to the methods for parameter estimation and the estimation of the parameter uncertainties (section \ref{sec-unc-params}).
Exploring the parameter space of the posterior pdf via MCMC sampling and comparing the mean and the MAP solutions and their uncertainties is recommended as a first validation step.
Multi-modal pdfs with similar weights around the posterior maxima cannot easily be summarized by single estimates and uncertainties.
Furthermore, they can result in misleading estimates in parameter ranges not supported by any of the diagnostics data \cite{DOSE99a}.
Such a multi-modal posterior pdf can be obtained, e.g., for inconsistent diagnostics data when the individual analyses of the diagnostics result in distant estimates with non-overlapping uncertainties, for outliers within a set of data, or for a misspecified uncertainty level for the data \cite{DOSE99a}.
As outliers of known or unknown source and misspecified uncertainties can be tackled with an outlier robust likelihood (section \ref{sec-likelihoods}), inconsistent diagnostics need to be studied in more detail. 
After a thorough inspection of sources for the inconsistency, candidate sources can be quantified with additional {\em nuisance} parameters, its uncertainties with a corresponding prior pdf, and marginalized (integrated out) from the posterior pdf.
This way the uncertainty of the nuisance parameter propagates to the parameters of interest.
If the posterior pdf becomes a unimodal distribution with parameter estimates and uncertainties capable of describing all diagnostics, a reasonable candidate for the inconsistency is found.
This way various candidates for the inconsistency can be tested and compared for their success in explaining all data simultaneously.
Please note, that with this method reasonable candidates can be identified, but the method does not guarantee to find the correct source of the inconsistency.
Nevertheless, this probabilistic method puts any inconsistency study on quantitative grounds.

Similar methods as for inconsistent diagnostics are applied for diagnostics deterioration, e.g., degradation of optical components (see also section \ref{sec-priors}).
An IDA method combining data from multiple diagnostics including calibration data provides a self-consistent approach to constrain uncertain and varying calibration {\em nuisance} parameters.
This approach becomes important in any harsh environment of future fusion devices as DEMO.

An extension of the validation methods described so far is given by the combination of measured data with modelling information.
Physical modelling allows to avoid non-physical prior information, to reduce the parameter space on physical grounds, and to validate the measured data.
Transport analyses, e.g., given by ASTRA modelling, might help to identify diagnostics lack of strength, e.g., unresolved parameter dependencies, as well as, e.g., limiting profile gradients not restricted by diagnostics data within their uncertainty. 
Validation typically performs best if all relevant information, measured data and modelling information, are jointly analysed.
The most important criterion is given by reasonable data residuals.
Successful validation needs data residuals within the uncertainties and data residuals scattering according to the likelihood pdf used.
Again, successful probabilistic validation does not imply a physically correct description of the data and correct physical modelling, but it provides a quantitative framework for the validation process.

\subsection{Numerical implementation}
\label{sec-num-impl}

Nowadays ample experience exists from applying IDA at various fusion devices (W7-AS, ASDEX Upgrade, JET, W7-X, TJ-II, and MST RFP), for various diagnostic combinations and for various parameter sets. 
Based on this experience and due to the conceptional clearly defined Bayesian approach an open-source IDA toolbox written in python and designed to be modular and flexible to be used at present and next generation fusion devices is presently under development.
The code is intended to be highly modular in the set of diagnostics considered, in the type of likelihoods to address different uncertainty conditions, in the multi-fidelity forward models (synthetic diagnostics) to allow for fast analysis with reduced physics for real-time applications up to post-plasma data analysis with highly-sophisticated diagnostics models, modular in the parameterisation (splines, Gaussian process regression, …), in the priors encompassing non-physical conditions (e.g. smoothness) or physical information from modelling codes, and modular in the evaluation and representation of results using MAP solutions or using MCMC sampling to explore the full probabilistic parameter space.  
A first implementation showing the combination of a synthetic set of interferometry, Thomson scattering and ECE diagnostics for the estimation of electron density and temperature profiles is described in section \ref{sec-prof-recon}. 

The IDA workflow is controlled by code parameters for, e.g., the selection of the set of diagnostics to be analysed, the likelihood and forward model to be used for each diagnostic, the equilibrium to be used, the time frame and temporal resolution with which the physical quantities are to be estimated, the parameterization of the physical quantities (profiles) to be estimated, the prior constraints to be applied (smoothness, physical models), or the parameter and uncertainty estimation methods (MAP, MCMC).
A frequently used format for code parameters is given by the XML format.
The present IDA implementation relies on the YAML format which is easier to be read and edited by humans.

\section{IMAS}
\label{sec-imas}

A multiple purpose data analysis framework should be adaptable to handle any data input and output method.
Nevertheless, a standardized communication scheme between codes and databases is beneficial for an efficient workflow.

The ITER Integrated Modelling \& Analysis Suite (IMAS) is the implementation of a physics modelling and data analysis suite for plasma operation and research. 
It provides standardized access to experimental and simulated data \cite{IMBEAUX15a}.
The data are organized in Interface Data Structures (IDS) which are designed for high modularity and flexibility to be suitable for any fusion device.
The IDS provide within a data dictionary a definition of data structures in a tree configuration and the names of the data to be used with the most popular programming languages.
New IDS are continuously developed and existing ones are extended according to the needs of code developers and users.

IMAS is designed to provide workflows for plasma modelling, data analysis and data structures.
The IDS encompass, e.g., the full description of the tokamak subsystems 
(diagnostic, heating system, etc.) or the physical concepts describing the plasma (equilibrium, set of core 
plasma profiles, wave propagation, etc.) \cite{IMBEAUX15a}. 
IMAS is used within the IDA framework, at present, by reading the machine description data for the diagnostics properties (geometry), the diagnostics data, diagnostics forward models, and the equilibrium. 
The linkage with IMAS will be extended as further diagnostics and forward models will be provided.

For a diagnostic forward model provided by IMAS to be used, the parameter representation internal to the IDA framework has to be mapped to the corresponding IDS needed as input to the IMAS forward model.
For example the spline representation of profiles within the IDA framework has to be mapped to the core-profile IDS defining the interface to the IMAS routines.

Eventually, the results of the data analysis, e.g., the estimated profiles and their uncertainties, and the forward modelled data and the residuals have to be written into the corresponding IMAS database for further usage.
The residuals, which describe the misfit of the measured data or modelling prior information with the forward modelled data, is of central importance for a (later) validation of the estimation results. 

Forward models provided by IMAS ideally have to be provided in a numerically-efficient modular way since the estimation of physical parameters requires a fitting (MAP solution) or a sampling (MCMC approach) procedure where the forward model is evaluated multiple times.
Separate sub-functions for initialisation and evaluation of synthetic signals, such that only the evaluation method is iterated in the IDA loop, have to be distinguished.
Three instances of the forward model are to be separated:
First, the initialization of time independent (static) quantities such as reading the geometry of the interferometer LOSs from the IMAS machine database which has to be done only once for the complete evaluation of a plasma pulse.
Second, the initialization of time dependent (dynamic) quantities such as the magnetic equilibrium and the magnetic coordinates along the interferometer LOSs, which has to be done once for each time point.
Third, the evaluation of the forward model (synthetic diagnostic signal) from the parameters to be estimated.
The third and innermost part of the IDA iteration loop defines the most critical part for numerical efficiency.

\section{Examples}
\label{sec-example}

\subsection{Synergistic effect}
\label{sec-synergy}

\begin{figure}[btp!]
\centering
\includegraphics[clip,width=80mm]{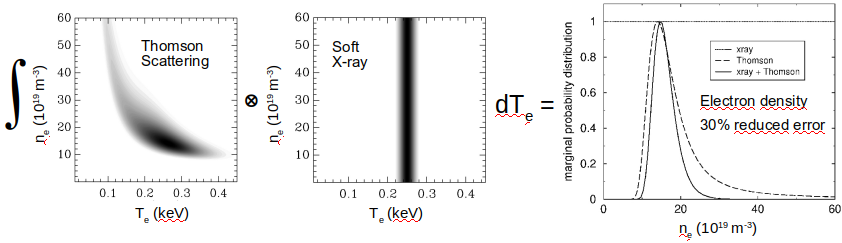}
\caption{Synergistic effect by exploiting full probability distribution}
\label{fig:synergy}
\end{figure}

The result of a Bayesian analysis is a probability distribution of the parameters of interest.
In case of a multidimensional probability distribution, the pdf contains the dependencies between the parameters.
These dependencies allow one to obtain a synergistic effect where the result of a diagnostic set is more informative than the sum of the results of individual measurements.
This is depicted in Fig.\ \ref{fig:synergy} where Thomson scattering data are analyzed together with soft X-ray data \cite{FISCHER03b}.
The left panel shows the 2-dimensional likelihood pdf as a function of density and temperature using only the TS data.
The hyperbolic structure is typical for this TS diagnostics which was most sensitive to the electron pressure.
The middle panel shows the pdf of a soft X-ray analysis where only temperature information was obtained, to be seen in the structure-less shape with respect to the density. 
Assuming we are interested in the density only, the 2-dimensional posterior pdfs have to be marginalized over the temperature.
The result of this marginalization is shown in the right panel for the Thomson data only (dashed curve), soft X-ray data only (dotted flat curve) and the joint analysis taken from the product of the two pdfs (solid line).
Although the soft X-ray data do not provide any information about the density, the joint analysis shows a 30\% reduction of the estimation uncertainty (width of the marginal distribution) of the density.
This example shows on the one hand the mechanism how a probabilistic synergistic effect is obtained, and, on the other hand, that exploiting the dependencies between the parameters are valuable for the combined analysis of heterogeneous diagnostics.

\subsection{Profile reconstruction}
\label{sec-prof-recon}

Various applications for profile reconstruction using IDA at the W7-AS stellarator \cite{FISCHER03b}, at the ASDEX Upgrade tokamak \cite{FISCHER08a,FISCHER10a,RATHGEBER10a,FISCHER20a}, at the JET tokamak \cite{FORD09a}, at the TJ-II stellarator \cite{FISCHER11a}, at the Madison Symmetric Torus (MST) reversed field pinch (RFP) \cite{REUSCH14a,GALANTE15a}, and at W7-X stellarator \cite{KWAK21a,KWAK21b} can be found.

Due to the conceptional clearly arranged Bayesian approach a general-purpose IDA toolbox, written in python, for present and next generation fusion devices was developed and will continuously be complemented as new diagnostics or parameterizations are requested.
The ingredients are summarized in chapter \ref{sec-IDA}.
A first application of this IDA toolbox implemented at ITER combines synthetic diagnostic data from artificial ECE and Thomson scattering (TS) diagnostics.
These two diagnostics were augmented with a synthetic data set from the Toroidal Interferometer Polarimeter (TIP) \cite{VANZEELAND13a}.
The IDA software package allows for selecting (via the YAML parameter file) for the diagnostics individually among Gaussian and Student's t-likelihoods and for the profile parameterization between a spline representation optionally with non-negativity constraints or an exponential of a spline representation which is by definition positive.
The profiles can be estimated in two ways, by finding the MAP solution of the posterior pdf or by evaluating the mean profiles from MCMC samples from the posterior pdf. 
Both estimation techniques allow to evaluate profile uncertainties.

\begin{figure}[tbp!]
\centering
\includegraphics[clip,width=100mm]{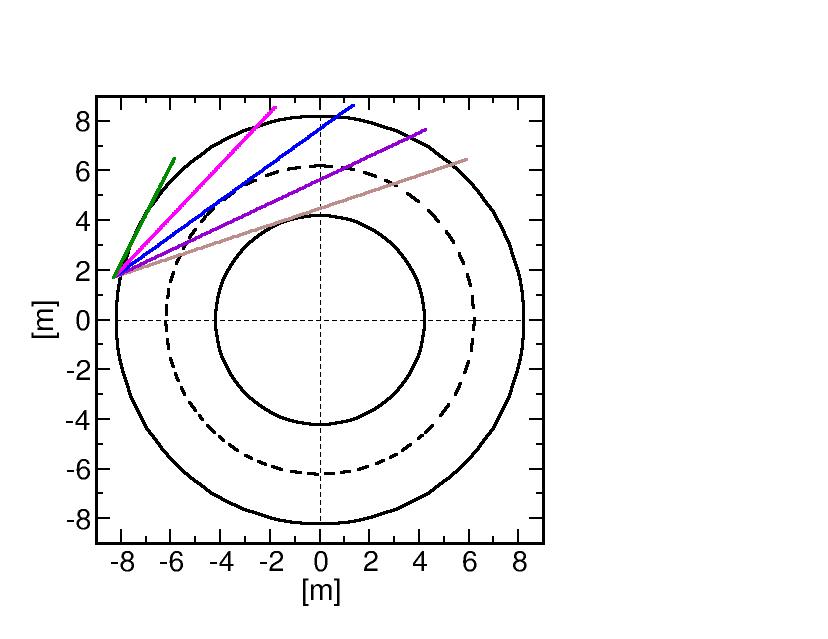}
\caption{Toroidal plane of ITER (major radius 6.2~m, minor radius 2~m) and the 5 TIP LOSs} 
\label{fig:tip_geom}
\end{figure}

The IDA software package reads from the ITER:IMAS database:
From the ITER machine description database the interferometry geometry of 5 lines of sight (Fig.\ \ref{fig:tip_geom}) and from the ITER scenario database an ITER equilibrium were taken.
For the TIP a synthetic data set was generated by line-integrating a core density profile corresponding to profiles from the ITER scenario database.
Random noise with a standard deviation of 5\% was added to the TIP data.
The ECE data and TS data were generated similarly at arbitrary positions within the plasma due to, at present, lack of realistic coordinates of the two diagnostics in the machine database. 
For the ECE forward model the basic assumption of a thermal and optically-thick plasma (black-body radiation) is assumed where the radiation temperature equals the electron temperature ($T_\textnormal{rad}=T_\textnormal{e}$) at the cold resonance position. 
This frequently used, trivial ECE forward model will be complemented with the radiation transport forward modelling using the ECRad code \cite{DENK18a,DENK20a}.
For simulating TS data, $T_\textnormal{e}$ and $n_\textnormal{e}$ pair values are taken at various positions in the plasma from the temperature and density profiles.
Random noise of 10\% for both ECE and TS data were added.

\begin{figure}[tbp!]
\centering
\includegraphics[clip,width=90mm]{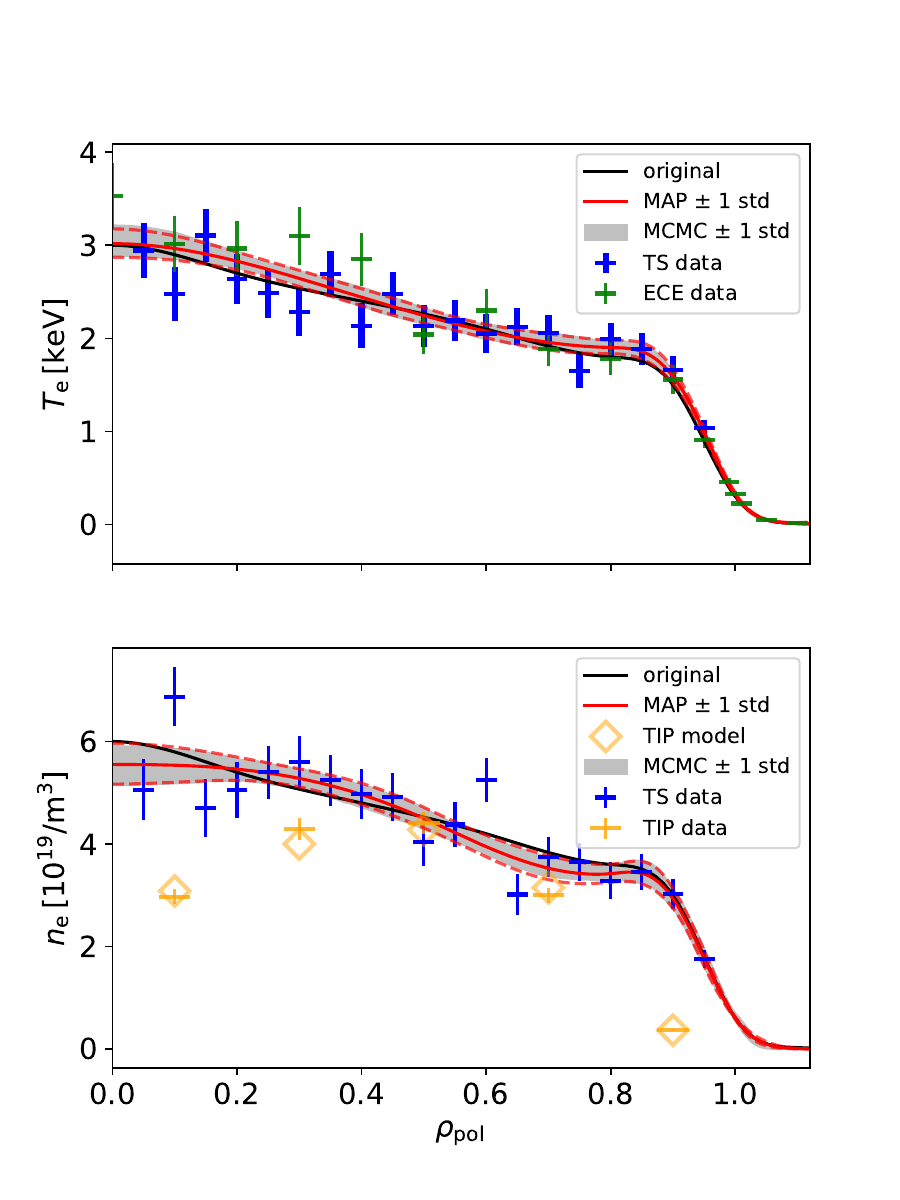}
\caption{Simulated (black solid line) and reconstructed (red solid line) temperature and density profiles estimated from the noisy data from the ECE (green), TS (blue) and TIP (orange) diagnostics. The open diamonds depict the forward modelled TIP data using the fitted density profile. The dashed area covers the uncertainty band from a MCMC sampling method.} 
\label{fig:tene_fit}
\end{figure}

The profiles taken for generating the data sets for the three diagnostics are shown in figure \ref{fig:tene_fit} as black solid lines (original).
The noisy data are shown as crosses (TS blue, ECE green, TIP orange) where the length of the vertical line corresponds to the uncertainty chosen.
As the TIP data are line integrated, the data are normalized to the lengths of the LOSs as shown in figure \ref{fig:tip_geom}.
The 5 TIP data are plotted at arbitrary plasma position sorted according to the smallest (largest) length of the LOS to the largest (smallest) $\rho_\mathrm{pol}$, correspondingly.

The profiles were parameterized with the exponential of a spline which ensures non-negativity.
For the ECE and TS data the Student's t and for the TIP data the Gaussian likelihood were arbitrarily chosen.
If sporadic fringe jumps in the TIP data are expected and routine (unsupervised) analysis is forseen, the Student's t-likelihood is beneficial in down-weighting the corrupted data, as routinely used within the IDA approach at ASDEX Upgrade.
No smoothing prior is applied.

The profiles estimated from the set of noisy data using the MAP solution are shown as red solid lines (MAP) with uncertainties employing the Gaussian approximation as red dashed lines. 
The profiles from the mean of the MCMC samples hardly deviates from the MAP profiles (not shown).
The uncertainty band estimated from the upper and lower standard deviations of the MCMC samples relative to the mean values are shown as shaded area (MCMC).
Please note that the use of upper and lower standard deviations typically result in asymmetric uncertainty bands if the posterior pdf is not symmetric with respect to its mean value. 
As expected, the MAP and MCMC estimates for this test example agree within the evaluated uncertainties with the original profiles.

\subsection{Equilibrium reconstruction}
\label{sec-equi-recon}

A reliable magnetic equilibrium reconstruction is essential for stability and transport studies as well as for the development of advanced plasma operation or steady-state tokamak operation with high bootstrap current fraction and non-inductive current drive \cite{BOCK17a,BOCK18a}.
Additionally, a reliable equilibrium is of major importance for the mapping of the diagnostics on a common coordinate system in the IDA framework.
Various equilibrium reconstruction codes and methods are available at the various fusion devices (see e.g.\ \cite{MCCARTHY99a,MCCARTHY12a,LAO05a,BRIX08a,BLUM12a,FISCHER16a}).
Frequently, for early availability and robustness equilibrium reconstruction is based on a reduced data set as, e.g., magnetic probe measurements only. 
But this usually results in reduced reliability especially of the core parameters (current and $q$-profiles and flux surfaces).
Therefore, for best performance the equilibrium reconstruction is part of the IDA workflow where a lot of relevant information for an improved equilibrium can be provided.

Often abundant measurement and modelling information is available 
for an improved reconstruction of the magnetic equilibrium.
This is exemplified with the ASDEX Upgrade equilibrium reconstruction using the IDE code package \cite{FISCHER16a,FISCHER19a}.
This framework is based on coupling of a Grad-Shafranov solver with current diffusion modelling.
The neo-classic current diffusion model (CDM) describes the temporal evolution of the equilibrium between two successive equilibrium reconstructions employing the Grad-Shafranov solver \cite{RAMPP16a}.
The CDM predicts the flux-surface averaged current density profile which provides data including their  uncertainties additionally to all the other measurements to constrain the next equilibrium reconstruction.
The free-boundary equilibrium solver employs data from magnetic measurements (field probes and flux measurements), diamagnetic measurements \cite{GIANNONE21a}, pressure profiles from electron  \cite{FISCHER10a} and ion temperature and density measurements \cite{FISCHER20a} and fast-ion pressure modelling (RABBIT \cite{WEILAND18a} for NBI or TRANSP \cite{BUDNY92a} for NBI and ICRH), effective ion charge $Z_{\rm eff}$ \cite{RATHGEBER10a}, internal current measurements from MSE and IMSE \cite{BURCKHART15a} and polarimetry \cite{MLYNEK16a}, tile (halo) currents for SOL currents, loop voltage measurements, $q$-constraints from mode analyses, topological iso-flux constraints from multiple measurements of $T_\textnormal{e}$ or $T_\textnormal{i}$ on the same flux surface \cite{LAO05a,FISCHER13a}, and plasma rotation measurements for considering centrifugal effects in an extended Grad-Shafranov equation \cite{LAO05a,FISCHER15a}.
For the neo-classical current diffusion the electron and ion temperature and density and the Zeff profiles are needed for the bootstrap current and the conductivity. 
Additionally, the electron cyclotron and neutral beam driven currents are provided by the TORBEAM and RABBIT codes.
Sawtooth reconnection events are described by two different sawtooth models for the sawtooth induced current re-distribution \cite{FISCHER19a}.
All these inputs provide redundant and complementary data for an improved and validated magnetic equilibrium.

The close interdependencies between the IDA profile estimation and the equilibrium reconstruction mutually influence also their reliabilities \cite{BRIX08a,FISCHER20a}.
As a fully integrated IDA approach covering profile estimation and equilibrium reconstruction simultaneously is still to be provided, a pragmatic approach is given by an alternating iteration of profile estimation and equilibrium reconstruction which was observed to converge within a few iterations.
The uncertainties of the input data for the equilibrium reconstruction are taken into account in the fitting part of the Picard iteration, and, therefore, propagate to the uncertainties of the equilibrium quantities.
The equilibrium uncertainties for the profile estimation can be considered by a Monte-Carlo approach sampling the base-function equilibrium coefficients from their covariance matrix and evaluate a random sample of equilibria to be used for the study of equilibrium-induced profile uncertainties.

\subsection{Integrated data analysis by velocity-space tomography}
\label{sec-tomo}

An application of integrated data analysis, which has emerged in recent years, is the measurement of fast ion velocity distribution functions by velocity-space tomography \cite{Salewski2012, Salewski2013, Salewski2014a}. As for any tomography application, integrated data analysis of all available measurements is essential. Position-space plasma tomography systems are usually designed with nominally identical or at least similar detectors. Velocity-space tomography uses any available detector monitoring the same spatial measurement volume, regardless of the type of diagnostic \cite{Salewski2013}. 

\begin{figure}[tbp!]
\centering
\includegraphics[clip,width=80mm]{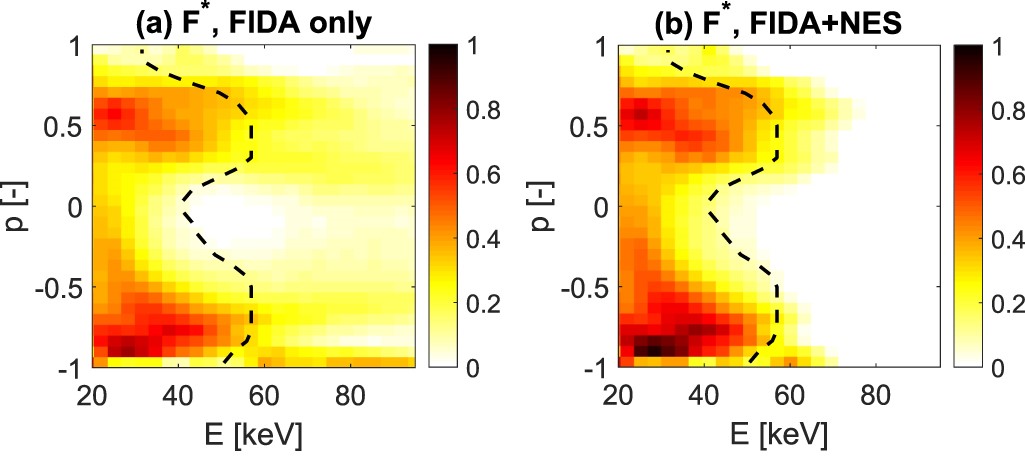}
\caption{Measurement of a fast-ion velocity distribution function [a.u.] in the center of a plasma heated by co-current and counter-current neutral beam injection at EAST \cite{Madsen2020a}. The tomographic inversion is based on (a) FIDA spectra using two detectors, and (b) additionally a NES spectrum. The expected ion densities to the right of the dashed line are low as expected from a calculation with TRANSP/NUBEAM.}
\label{fig:EASTtomo}
\end{figure}

An example appears in figure~\ref{fig:EASTtomo} showing the velocity distribution function at EAST for a plasma heated by co- and counter-current neutral beam injection (NBI) \cite{Madsen2020a}. $E$ is the energy and $p$ is the pitch. The measurements are analyzed using (a) only two fast-ion D$_\alpha$ (FIDA) spectroscopy detectors and (b) the two FIDA detectors and in addition neutron emission spectroscopy (NES). 
The dashed line represents the upper boundary of a velocity distribution function computed with TRANSP/NUBEAM. Few ions are expected to the right of the dashed line because only NBI heating was used and no acceleration of ions is expected. If, on the one hand, only the two FIDA detectors are used to compute the tomographic image, large ion densities at energies significantly larger than the NBI injection energies are erroneously found. We additionally recognize these as artifacts since similar artifacts appear in tomographic inversions of synthetic measurements, where the true solution is known, for this diagnostic setup. If, on the other hand, in addition NES measurements are used, they force the distribution function to very small values at large energies. This works well since NES measurements are highly sensitive at large energies \cite{Moseev2018, Eriksson2019}. This example shows that the integrated data analysis of FIDA and NES here suppresses the artifacts at large energies. A second example in addition to figure~\ref{fig:EASTtomo} is velocity-space tomography based on $\gamma$-ray spectroscopy (GRS) measurements and NES measurements at JET \cite{Salewski2017}. The NES measurements are made by time-of-flight, liquid scintillator and single-crystal diamond detectors \cite{Eriksson2015}, so that in total four different detector types were used in the tomographic inversion.

Until now velocity-space tomography has been applied at ASDEX Upgrade \cite{Salewski2014a, Salewski2016b, Weiland2016, Weiland2017, Jacobsen2016a, Jacobsen2016b, Rasmussen2016}, JET \cite{Salewski2017}, MAST \cite{Madsen2018}, DIII-D \cite{Madsen2020a}, EAST \cite{Madsen2020b, Su2021} and TCV \cite{Geiger2017}. Various combinations of data from FIDA, NES, GRS as well as collective Thomson scattering (CTS) with two to five simultaneous detectors have been used at these tokamaks. Velocity distribution functions in plasmas heated by neutral beam injection (NBI) as well as electromagnetic wave heating in ion cyclotron range of frequencies (ICRF) have been measured. 

At ITER, velocity-space tomography of the $\alpha$-particle distribution function based on GRS and CTS has been shown to be feasible for energies from about 1.7~MeV upwards \cite{Salewski2018c}. However, since all currently foreseen diagnostics observe in a perpendicular direction with respect to the magnetic field, the sign of the pitch $p$ of the $\alpha$-particles cannot be determined. But the absolute value $|p|$ can be determined, so that the velocity distribution function $f(E,|p|)$ can be measured. If an oblique $\gamma$-ray detector is installed, the sign of the pitch can be found, too \cite{Salewski2018c}.

Velocity-space tomography is also applied to measure anisotropic deuterium temperatures $T_\|$ and $T_\perp$ as well as the deuterium density and drift velocity \cite{Salewski2018a}. In this  application, the full (fast and thermal) velocity distribution function is determined based on simultaneous measurements with several $D_\alpha$-spectroscopy detectors, and then the lowest moments of the full velocity distribution function are computed \cite{Salewski2018a}. 

Reviews of velocity-space tomography are available in references \cite{Moseev2018, Salewski2019dr}. In the following we will focus on methods of velocity-space tomography and discuss the forward model, the inverse problem, prior information, uncertainties as well as related 1D to 5D tomography problems.

\subsubsection{The forward model: Synthetic diagnostics}
\begin{figure}[tbp!]
\centering
\includegraphics[clip,width=80mm]{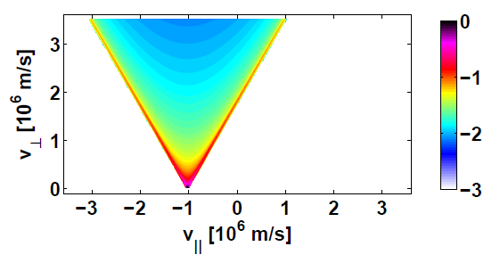}
\caption{Exemplary weight function showing the velocity-space sensitivity of a CTS measurement at a particular Doppler shift.}
\label{fig:wfCTS}
\end{figure}

To do tomography in velocity space, we need to quantify the sensitivity of the diagnostics in velocity space. This is analogous to modeling the lines-of-sight in traditional position-space tomography. Weight functions quantifying this velocity-space sensitivity have been developed for all major fast-ion diagnostics: FIDA \cite{Heidbrink2007, Salewski2014b}, neutral particle analyzers (NPA) \cite{Heidbrink2007}, CTS \cite{Salewski2011}, NES \cite{Jacobsen2015, Jacobsen2017}, GRS \cite{Salewski2015b, Salewski2016a} and fast-ion loss detectors \cite{Galdon-Quiroga2018b}. Recently, weight functions for 3~MeV proton diagnostics \cite{Heidbrink2021} and ion cyclotron emission spectroscopy weight functions \cite{Schmidt2021} have been numerically computed, too. A weight function $w$ relates a 2D fast-ion distribution function $f$ to a measurement $s$ through the integral equation \cite{Heidbrink2007, Salewski2011, Salewski2014b, Jacobsen2015,Jacobsen2017, Salewski2015b, Salewski2016a}
\begin{eqnarray}
\label{eq:wdefvpavpe}
s(m_1,m_2)&=&\int_0^\infty \int_{-\infty}^\infty w (m_1, m_2, v_\|, v_\perp) \nonumber \\ 
&\times& f(v_\| , v_\perp ) d v_\|  d v_\perp .
\end{eqnarray}
$s(m_1,m_2)$ is the measured signal in the spectral bin $[m_1,m_2]$. FIDA measures spectra in wavelength \cite{Heidbrink2010}, GRS in $\gamma$-ray energies \cite{Nocente2020}, time-of-flight NES in flight times \cite{Eriksson2019}, and CTS in wave frequency \cite{Salewski2010}. $(v_\|, v_\perp)$ are the velocities parallel and perpendicular to the magnetic field, respectively. ($E,p$) coordinates are also often used but in $(v_\|, v_\perp)$ the geometrical shape of weight functions is often clearest. The weight function hence shows the quantity [signal / fast ion] where the units of the signal are particular to the instrument. An example of a weight function appears in figure~\ref{fig:wfCTS}. The colored regions are observable for the given measurement bin whereas the white regions are not observable.

Substitution of a $\delta$-function modeling $N_f$ ions at coordinates ($v_{\| 0}, v_{\perp 0}$) into equation~\ref{eq:wdefvpavpe} and integration shows that the amplitude of a weight function at velocity-space position $(v_{\|0}, v_{\perp 0})$ is readily computed from 
\begin{eqnarray}
w (m_1, m_2, v_{\| 0}, v_{\perp 0})=\frac{s(m_1,m_2)}{N_\textrm{f}}
\label{eq:numericwfc}
\end{eqnarray}
using a standard synthetic diagnostic code for the diagnostic. The computation of the signal by weight functions neglects spatial effects, which is usually fairly accurate for the plasma center where plasma profiles are flat and spatial effects are hence negligible. 

Knowing the weight functions for all available measurement bins in a measured spectrum, we can write the forward model of the diagnostic as the matrix equation
\begin{eqnarray}
S=WF
\label{eq:SWF}
\end{eqnarray}
which summarizes a discretization of equation~\ref{eq:numericwfc} \cite{Salewski2012}. $S$ is a vector holding the measurement data of all available diagnostics, $F$ holds the velocity distribution function rearranged as column vector, and each line of the matrix $W$ holds a weight function rearranged as a row vector. Given a simulation $F$ and knowing the weight matrix $W$, we can rapidly compute the corresponding synthetic signals $S$ for all diagnostics.

\subsubsection{Likelihood}
\label{subsec-likelihood-tomo}
To determine $F$, given $W$ and $S$, equation~\ref{eq:SWF} has mathematically the same form as any traditional position-space tomography problem. However, velocity-space tomography often requires combinations of entirely different diagnostics or detector types. For practical reasons, to combine measurements with vastly different amplitudes by orders of magnitude, each individual measurement in $S$ and its corresponding weight function is normalized by its uncertainty \cite{Salewski2013}. 
This is equivalent to the unnormalized data if likelihoods with the $\chi^2$-misfit between the data and the forward modelled data weighted with the measurement uncertainty are used as, e.g., for the Gaussian and Student's t-likelihood (equations \ref{eq:gauss} and \ref{eq:cauchy}). In the present inverse problem of velocity-space tomography the Gaussian likelihood is employed.

\subsubsection{Prior information}
\label{subsec-prior-tomo}
The amount of measured fast-ion diagnostic data is always small in fusion plasmas due to the limited access to the plasma and the often comparatively low signal-to-noise ratio for tomography applications. Furthermore, there are never more than just a few detectors, such that we must always determine the 2D image from just a few projections (medical tomography uses hundreds of projection directions).
Due to this limited amount of data and projection directions, the use of prior information for this ill-posed inference problem is often essential to reduce noise fitting and obtain meaningful images \cite{Salewski2016b}, in particular in velocity-space regions observed by only one or two detectors \cite{Salewski2018b, Salewski2018c, Madsen2018, Madsen2020a}. 

As in many tomography problems, the problem to find $F$ from $W$ and $S$ is ill-posed and must be regularized with additional (prior) information. The most successful regularization method in velocity-space tomography has been the Tikhonov regularization in which we solve a least-square problem of the form \cite{Salewski2016b}
\begin{eqnarray}
\underset{F}{\text{minimize}} \Bigg\| \Bigg(\begin{array}{c}W\\ \lambda L\end{array}\Bigg) F- \Bigg(\begin{array}{c}S\\ 0\end{array}\Bigg) \Bigg\|_2
\label{eq:Tikhonov}
\end{eqnarray}
The upper row of this matrix equation seeks to fit the data whereas the lower row penalizes undesired features of the solution. $L$ is the regularization matrix. Velocity-space tomography usually uses zeroth order Tikhonov regularization, where $L$ is the identity matrix, or first-order Tikhonov regularization, where $L$ is a matrix effecting a gradient. This penalizes large absolute values or large gradients, reflecting our prior information that we believe the velocity distribution function to be smooth due to collisions. Equation \ref{eq:Tikhonov} is equivalent to maximizing a posterior pdf given by the product of a Gaussian likelihood with a Gaussian prior pdf with the Tikhonov term in the exponent \cite{Salewski2018b}. 

$\lambda$ is the regularization strength that balances data fitting versus the regularization requirement. $\lambda$ is a free parameter of the problem that must be determined as part of the solution. Various methods to choose $\lambda$ automatically have been tested, e.g. the L-curve or the generalized cross validation method \cite{Salewski2016b}. However, no method is clearly always advantageous, and they usually produce similar $\lambda$'s within a factor 10. It is advisable to inspect a range of $\lambda$'s to make sure that no phenomena are missed. 

This is the standard regularization technique in many plasma tomography applications. If no constraints are introduced, the solution is given by the so-called normal equation
\begin{eqnarray}
F_\lambda=\big(W^T W + \lambda^2 L^T L \big)^{\scalebox{0.75}[1.0]{-} 1}W^T S.
\end{eqnarray}
We write the index $\lambda$ in $F_\lambda$ to emphasize that the solution depends on the regularization strength. However, $F_\lambda$ computed with the normal equation usually becomes negative in some velocity-space regions, which is unphysical. This can be remedied by further prior information about non-negativity constraints.



We are certain that the fast-ion velocity distribution function is not negative. This prior can be encoded by solving a least-square problem with non-negativity constraint  \cite{Salewski2016b}:
\begin{eqnarray}
\underset{F}{\text{minimize}} \Bigg\| \Bigg(\begin{array}{c}W\\ \lambda L\end{array}\Bigg) F- \Bigg(\begin{array}{c}S\\ 0\end{array}\Bigg) \Bigg\|_2 \hspace{0.3cm} \textnormal{subject to}\hspace{0.2cm}  F \ge 0. \nonumber \\
\label{eq:TikhonovNonNeg}
\end{eqnarray}
One can simply use a non-negative least-squares algorithm \cite{LawsonHanson1974}. Alternatively, one may penalize negative values and hence force them to become small \cite{Weiland2016}. The non-negativity constraint also acts as a useful smoothing regularizer since it tends to dampen high-frequency spatial oscillations in the solution. Since the prior information of non-negativity is absolutely certain, we regard the non-negative Tikhonov problem as our standard method.

\begin{figure}[tbp!]
\centering
\includegraphics[clip,width=80mm]{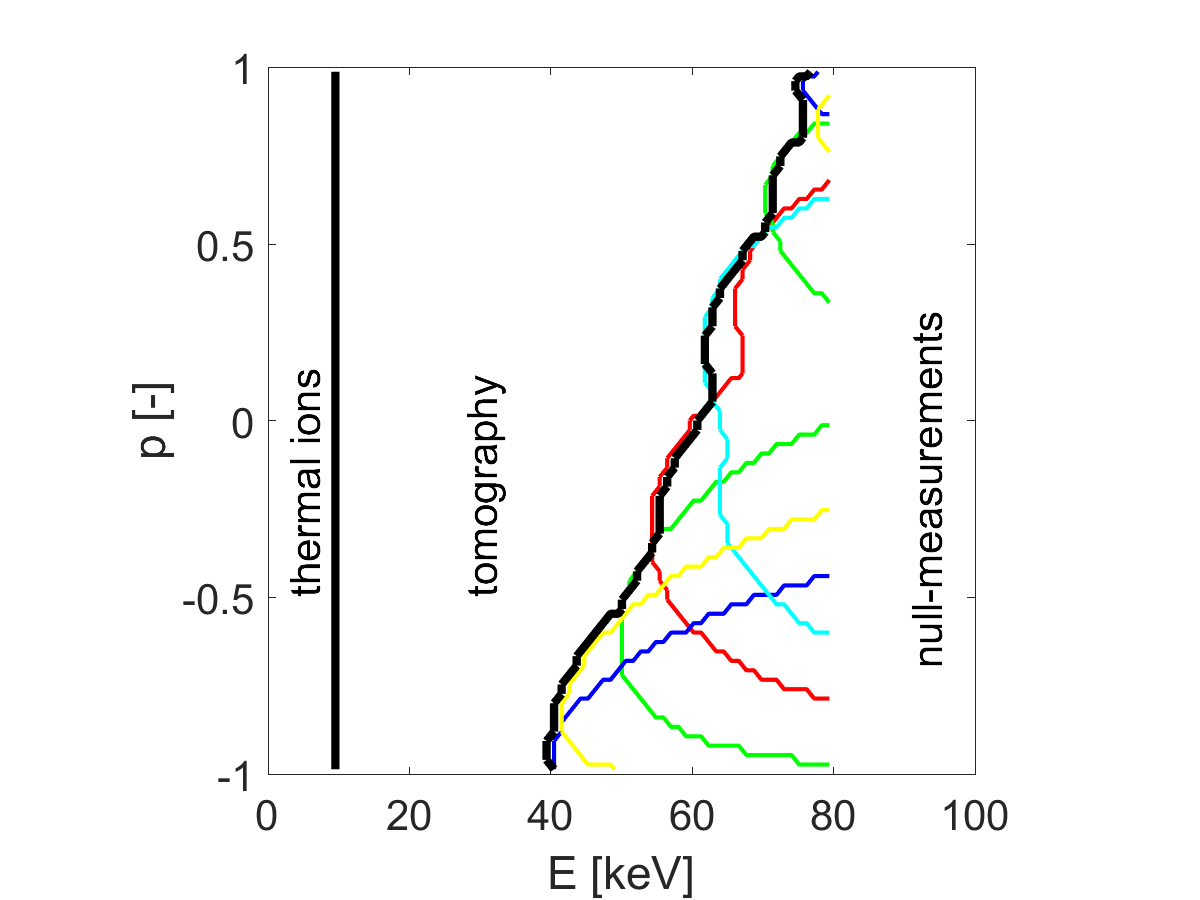}
\caption{The colored lines are boundaries of weight functions connected to null-measurements. The black line is their envelope, presenting a boundary to the velocity space region empty of fast ions \cite{Salewski2016b}.}
\label{fig:nullmeasurments}
\end{figure}

Several other constraints have been implemented in addition to non-negativity: isotropy, monotonicity, or restrictions on the target velocity space to be reconstructed. 
For example, a minimization problem with non-negativity, restricted velocity space, and monotonicity constraint in energy can be written as
\begin{eqnarray}
\underset{F}{\text{minimize}} \Bigg\| \Bigg(\begin{array}{c}W\\ \lambda  L\end{array}\Bigg) F- \Bigg(\begin{array}{c}S\\ 0\end{array}\Bigg) \Bigg\|_2 \hspace{0.3cm} \nonumber \\ \textnormal{subject to}
\begin{cases}
F \ge 0, \\ F(E_0, p_0)=0, \\ L_{1,E} F (E_\textrm{m}, p_\textrm{m})\leq 0.
\end{cases}
\label{eq:TikonovNullConstraints}
\end{eqnarray}
$F(E_0,p_0)=0$ is the velocity-space region with negligible fast-ion densities according to null-measurements, as identified by weight functions \cite{Salewski2016b}. A null-measurement in the measured signal $S$ is the measured absence of evidence for fast ions. An example of an experimentally determined null-measurement velocity space region at ASDEX Upgrade is illustrated in figure~\ref{fig:nullmeasurments}.  It is advantageous to use null-measurements as they restrict the velocity space by reducing the number of unknowns \cite{Salewski2016b}. Null-measurements are perhaps more intuitively understood in position-space tomography problems: A ray that misses the object altogether will measure the absence of any material, and thus this part of position-space does not need to be reconstructed.

A monotonicity constraint in one of the coordinate directions, in equation~\ref{eq:TikonovNullConstraints} the energy, can be advantageous if one is confident that the distribution function is monotonous \cite{Madsen2020a}. This is likely a good assumption for $\alpha$-particles or usually ICRF fast-ion tails. However, any local minimum in the distribution function may be physical, which would be missed when this mathematical constraint is enforced. 

\begin{figure}[tbp!]
\centering
\includegraphics[clip,width=57mm]{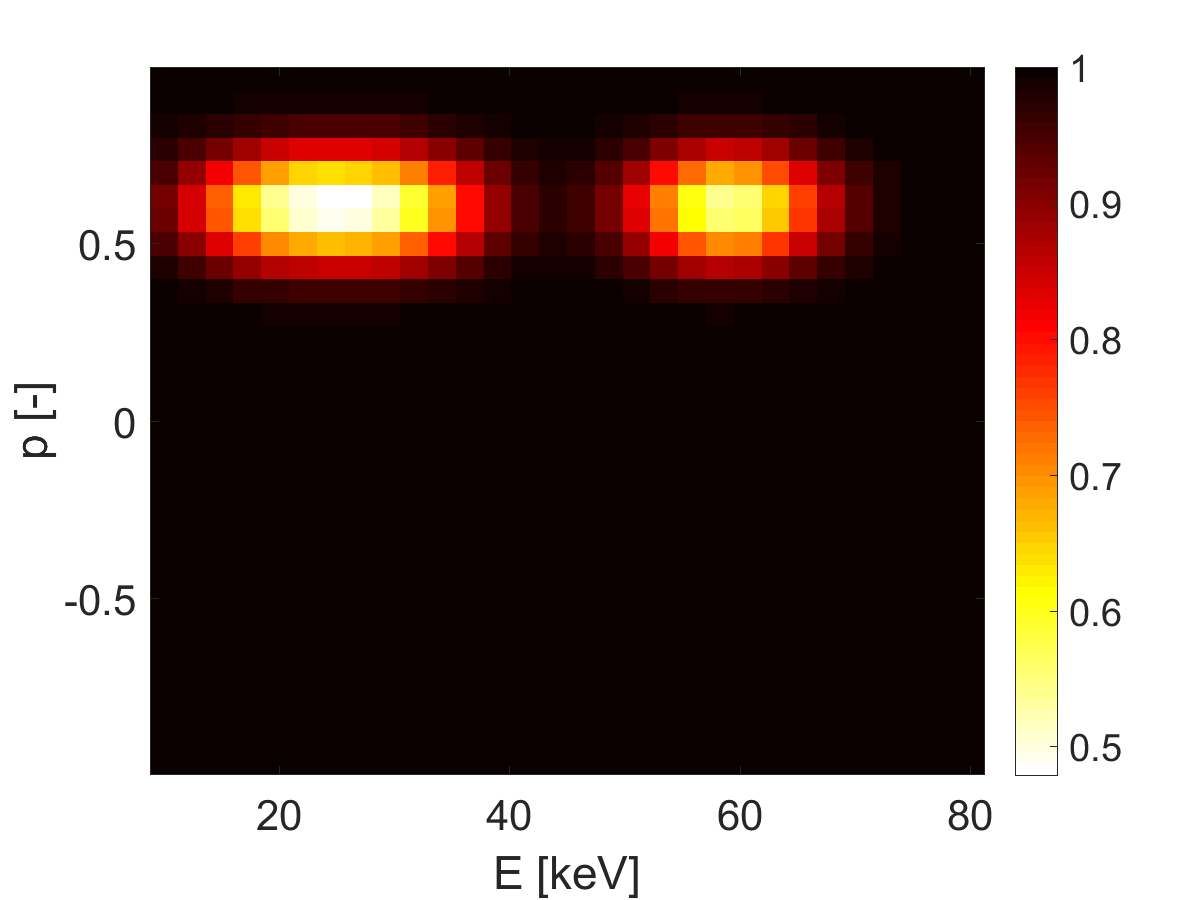}
\caption{$\kappa_1(E,p)$ encodes the NBI injection energies and pitch.}
\label{fig:lambda2D}
\end{figure}

Prior information may also be encoded by modifying the penalty function to become dependent on the velocity-space coordinates. For example, to promote nearly isotropic solutions, we can penalize gradients in pitch direction much more strongly than in energy direction \cite{Salewski2018b, Salewski2019}. This idea is similar to anisotropic regularization along flux surfaces in position-space plasma tomography. We may further enforce isotropy as constraint by assuming the solution to be constant in pitch direction \cite{Salewski2019dr}.

Another way to modify the penalty function is to introduce a function $\kappa_1 (E,p)$ acting with a first-order Tikhonov regularizer or another function $\kappa_0 (E,p)$  acting with a zeroth-order Tikhonov regularizer \cite{Salewski2016b, Madsen2018, Madsen2020a, Madsen2020b}. The minimization problem with a mix of zeroth- and first-order Tikhonov terms is written as
\begin{eqnarray}
\underset{F}{\text{minimize}}  \left\| \left( \begin{array}{c} W \\ \lambda_1 \kappa_1(E,p) L_1 \\ \lambda_0 \kappa_0(E,p) L_0 \end{array} \right) F - \left( \begin{array}{c} S \\ 0 \\ 0 \end{array} \right) \right\|_2.
\label{eq:TikhonovSoftNull}
\end{eqnarray}
$\kappa_1(E,p)$ used with the first-order Tikhonov penalty term $\lambda_1 L_1$ can encode the velocity-space positions of the particle sources of an NBI. The velocity-space positions of the particle sources at the full-, half- and one-third NBI injection energies at a particular pitch are well-known. If $\kappa_1(E,p)$ is chosen to have minima at these well-known peak locations, as illustrated in figure~\ref{fig:lambda2D}, gradients are penalized less in the vicinity of the particle sources than elsewhere \cite{Salewski2016b}. This can allow the formation of peaks in the image but does not force it. Observe that when the gradients are penalized less, a local minimum is equally well formed as a local maximum. Data will usually dictate the formation of a maximum. 

\begin{figure}[tbp!]
\centering
\includegraphics[clip,width=80mm]{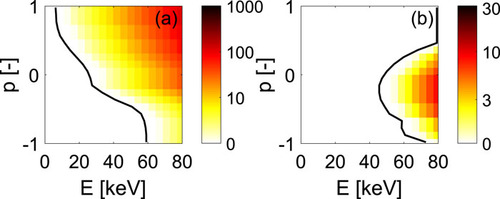}
\caption{Prior information of unlikely velocity space for velocity-space tomography at MAST according to (a) TRANSP/NUBEAM and (b) null-measurements \cite{Madsen2018}. The monotonously growing $\kappa_0(E,p)$ towards higher energies expresses our increasing doubt to find ions.}
\label{fig:lambda2Dnull}
\end{figure}

Instead of a formulation of null-measurements as a mathematical constraint, we can introduce a zeroth order Tikhonov penalty in the null-measurement velocity space as $\lambda_0 \kappa_0(E,p)L_0$ \cite{Madsen2018}. This method can be used if we are in doubt if the velocity-space is empty, e.g. if the null-measurements are too uncertain to set the related velocity space to zero. An example of a function $\kappa_0$ for a velocity-space tomography problem at the MAST tokamak appears in figure~\ref{fig:lambda2Dnull}.

The increasing penalty with energy reflects our increasing doubt to find ions at increasing energies \cite{Madsen2018}. If the boundary between the null-measurement velocity space and the target velocity space cannot be found from measurements, we can use a numerical simulation, for example using TRANSP/NUBEAM, to restrict the velocity-space region considered for the inversion \cite{Madsen2018}. This assumes that there is no acceleration mechanism for fast ions to accelerate them beyond the boundary from the neoclassical simulation.

A recent idea is to expand the velocity distribution function into a series of slowing-down distribution functions $f_{\textrm{SD}}$ \cite{Madsen2020b}. The tomography problem is then to determine the coefficients $a_i$ associated with the $i$th base function. In matrix form this expansion can be written as
\begin{equation}
F=F_{\textrm{SD}}A
\end{equation}
where the vector $A$ holds the expansion coefficients,  and the matrix $F_{\textrm{SD}}$ holds the slowing-down distribution functions rearranged as columns. The forward problem becomes
\begin{equation}
S=WF_{\textrm{SD}}A.
\end{equation}
Given $A$ and knowing $W$ and $F_{\textrm{SD}}$, we can calculate the signal $S$. Calculating $A$, given $S$, is an ill-posed problem, as the tomography problem with the expansion. We can then solve a zeroth-order Tikhonov problem in the expansion coefficients $A$ of the form
\begin{eqnarray}
\underset{A}{\text{minimize}} \Bigg\| \Bigg(\begin{array}{c}WF_{\textrm{SD}}\\ \lambda I\end{array}\Bigg) A- \Bigg(\begin{array}{c}S\\ 0\end{array}\Bigg) \Bigg\|_2.
\label{eq:TikhonovSDexpansion}
\end{eqnarray}
If $F_{\textrm{SD}}$ is invertible, we can substitute $A=F_{\textrm{SD}}^{-1}F$, and equation~\ref{eq:TikhonovSDexpansion} can be equivalently written as
\begin{eqnarray}
\underset{F}{\text{minimize}} \Bigg\| \Bigg(\begin{array}{c}W\\ \lambda F_{\textrm{SD}}^{-1}\end{array}\Bigg) F- \Bigg(\begin{array}{c}S\\ 0\end{array}\Bigg) \Bigg\|_2
\label{eq:TikhonovSD}
\end{eqnarray}
which shows that the expansion in basis functions can be formulated as a standard Tikhonov problem with $F_\textrm{SD}^{-1}$ as the regularization matrix $L$. This allows us to interpret this expansion in slowing-down basis functions as slowing-down regularization. This type of regularization reflects our prior belief that the usual slowing-down physics will in part determine the shape of the distribution function. However, if data dictates otherwise, deviations will appear due to the upper row of equation~\ref{eq:TikhonovSD}.

Lastly, if there is not enough data to do a full tomographic inversion, we can use a numerical simulation $F_\textrm{sim}$ as prior information and penalize any deviation from the simulation \cite{Salewski2016b}. The penalty term becomes $\|L(F-F_\textrm{sim})\|_2$,
and we write the problem as
\begin{eqnarray}
\underset{F}{\text{minimize}} \hspace{0.2cm} \Bigg\| \Bigg(\begin{array}{c}W\\ \lambda  L\end{array}\Bigg) F- \Bigg(\begin{array}{c}S\\ \lambda LF_\textrm{sim}\end{array}\Bigg) \Bigg\|_2 
\label{eq:TikonovSimPrior}
\end{eqnarray}
Other prior information can still be added in the same way as described. However, observe that this Tikhonov problem pursues a less ambitious goal than a full tomographic inversion due to the use of the simulation. Our goal here is only to identify regions in velocity space where the measurements suggest deviations from $F_\textrm{sim}$ which we can find by successively decreasing $\lambda$. 

We summarize the different types of prior information that has be used in velocity-space tomography:
\renewcommand\labelitemi{\tiny$\bullet$}
\begin{itemize}
    \item smoothness (zeroth- or first-order Tikhonov)
    \item non-negativity constraint or penalty for negative values
    \item restricted velocity space by null-measurements or according to a simulation (constraint)
    \item unlikely velocity space by null-measurements or according to a simulations (penalty)
    \item monotonicity constraint
    \item isotropy constraint or penalty for deviation from isotropy
    \item beam injection peak locations
    \item slowing-down physics as regularizer
    \item numerical simulation as prior
\end{itemize}


\subsubsection{Uncertainties}
\label{subsec-unc-tomo}
Sources for uncertainties in the estimated velocity-space distribution can be divided into four categories: 1) uncertainties due to (statistical) measurement noise  \cite{Salewski2013}, 2) bias uncertainties in the measurements (systematic uncertainties), 3) uncertainties in the weight matrix $W$ (forward model) due to uncertainties in nuisance parameters \cite{Jacobsen2016b} and 4) bias uncertainties due to the regularization \cite{Jacobsen2016b} (prior  information). 

Analytic formulas for estimating uncertainties due to measurement noise and uncertainties in the nuisance parameters can be given for the unconstrained Tikhonov problem \cite{Jacobsen2016a, Salewski2018c}. For the constrained Tikhonov problem, these uncertainties can be found by sampling. The fast-ion measurements and the nuisance parameters are sampled from their probability distributions. For each sample we obtain an inversion. Hence we generate a population of $N$ inversions, $F_{\lambda,i}$. Its mean is the best estimate of the velocity-distribution function, and its standard deviation is the uncertainty due to uncertainties in the signal and the nuisance parameters:
\begin{eqnarray}
\langle F_\lambda \rangle &=&\frac{1}{N} \sum_i F_{\lambda,i}, \\
\delta F_\lambda &=& \sqrt{\frac{1}{N-1}\sum_i \left(F_{\lambda,i}- \langle F_\lambda \rangle\right)^2}.
\end{eqnarray}
Each pixel in the inversion hence has its own uncertainty \cite{Jacobsen2016a, Jacobsen2016b}. The two contributions can also be individually calculated if required.

The computation of bias uncertainties due to regularization and due to systematic errors in the measurements are open problems. Bias uncertainties make it impossible to reconstruct the true distribution function even with perfect, noise-free measurements. To quantify this bias, we would need to know the true solution, which we never do \cite{Salewski2018b}. 

Systematic bias uncertainties in the measurement data are also notoriously difficult to detect. But such systematic measurement errors can lead to systematic artifacts which can sometimes reveal that some error is present. For example, errors in the calibration of the measurements lead to systematic artifacts that can give clues to what might be wrong.

\subsubsection{Discussion}
To make optimal use of the diagnostic set at ITER or other tokamaks and stellarators, we must develop 1D to 5D inversion tools and be able to use prior information to deal with the sparsity of data on these devices. The methods presented here will allow measurements of $\alpha$-particle velocity distribution functions for energies from 1.7 MeV upwards based on IDA of GRS and CTS \cite{Salewski2018c}. Hence energy spectra in ITER, as requested by the ITER measurement requirements \cite{Donne2007}, can also be determined. Below $\alpha$-particle energies of 1.7~MeV, CTS will be the only diagnostic for $\alpha$-particles. If only one detector is available, 1D inversion techniques need to be used to determine energy spectra, for example by assuming isotropy or near-isotropy as prior \cite{Salewski2019, Salewski2019dr}.

It is now also becoming possible to measure 3D phase-space distribution functions by orbit tomography \cite{Stagner2017, Jaerleblad2021}. This approach is analogous to velocity-space tomography, but in 3D phase-space of constants of motion which covers the entire ion population in the tokamak. Each grid point in 3D phase space corresponds to an orbit in the tokamak.
The forward model computes the signal generated by fast ions on each orbit. In the tomography problem, the 3D phase-space distribution function of all fast ions in the plasma is inferred from the measurements of all detectors. The computed orbits constitute the prior information for orbit tomography. This approach has worked well at ASDEX Upgrade \cite{STAGNER22a} and is being implemented at JET \cite{Jaerleblad2021}. It requires many measurements to cover the 3D target phase space, but with appropriate additional prior information it will hopefully be useful at ITER.

Lastly, the presented method to expand the distribution function in 2D base functions is actually not restricted to 2D. The base functions can be 3D, 4D or 5D functions. For non-axisymmetric plasmas, the entire phase-space distribution function is described by 3D in position space and only 2D in velocity space due to the fast gyromotion. 5D tomography would allow integrated data analysis of all measurements on stellarators or non-axisymmetric tokamaks.

\section{Summary}
\label{sec-summary}

Integrated data analysis in the framework of Bayesian probability theory provides a method for improved results by a coherent combination of heterogeneous diagnostic measurements with physical prior and modelling information to restrict the parameter space of otherwise ill-posed inversion techniques.
The concept of IDA is outlined and contrasted with conventional data analysis.
The ingredients of this probabilistic approach is given by forward modelling, suitable likelihood pdfs with comprehensive uncertainty quantification of measurements, probabilistic quantification of prior information, and probabilistic quantification of nuisance parameters and their marginalization.  
The probabilistic approach enables us to obtain synergistic effects by exploiting the parameter correlation structure and diagnostics interdependencies.

A general purpose IDA toolbox was developed for present and next generation fusion devices and applied to a combination of ITER profile diagnostics.
Profile estimation and equilibrium reconstruction is closely correlated and is recommended to be combined in an IDA approach.
An example of IDA by velocity-space tomography highlights the benefit of combining various heterogeneous diagnostics with physical prior information including their uncertainties.

\section*{ACKNOWLEDGEMENT}

The authors thank S. Pinches from the ITER organization, K. Fujii from the Kyoto University and D. Mazon from CEA Cadarache for promoting and supporting this new working group.

This work has been carried out within the framework of the EUROfusion Consortium, funded by the European Union via the Euratom Research and Training Programme (Grant Agreement No 101052200 — EUROfusion). Views and opinions expressed are however those of the author(s) only and do not necessarily reflect those of the European Union or the European Commission. Neither the European Union nor the European Commission can be held responsible for them.

\section*{REFERENCES}

\newcommand{\newblock}{}
\bibliographystyle{unsrt}
\bibliography{fusion.bib,salewski2017bibshort.bib}
\end{document}